\documentclass[prb,
preprint,
showpacs,
amsmath,
amssymb,
floatfix]
{revtex4-1}

\usepackage{units}
\usepackage{psfrag}

\usepackage[dvips]{graphicx}

\newcommand{\e}{\varepsilon}              %

\begin{document}

\title{Robust synchronization of spin-torque oscillators with an LCR load}

\author{Arkady Pikovsky} 
\affiliation{Department of Physics and Astronomy, Potsdam University, 
  Karl-Liebknecht-Str 24, D-14476, Potsdam, Germany}
\affiliation{NEXT, Laboratoire de Physique Th́{\'e}orique du CNRS, IRSAMC, Universit{\'e} Toulouse, UPS, 31062 Toulouse, France}  

\date{\today}

\begin{abstract}
We study dynamics of a serial array of spin-torque-oscillators with a parallel 
inductor-capacitor-resistor (LCR) load. In a large range of parameters the fully synchronous regime, where
all the oscillators have the same state and the output field is maximal, is shown to be stable.
However, not always such a robust complete synchronization develops from the random initial state,
in many cases nontrivial clustering is observed, with a partial synchronization resulting in
a quasiperiodic or chaotic mean field dynamics.  
\end{abstract}


\maketitle

\section{Introduction}
Spin-torque oscillator (STO) is a nanoscale spintronic device generating periodic 
microwave (in the frequency range of several GHz) oscillations 
(see~\cite{Slavin-Tiberkevich-09} for an
introductory review). The physics behind 
these oscillations is based on the spin-transfer torque force, with which a spin-polarized 
electrical current acts on a free magnet. Sometimes one uses terms ``Spin-transfer oscillator'' 
or ``Spin-transfer nano-oscillator'' (STNO) to describe this object. 
The STO consists of two magnetic layers, one (bottom)
having fixed magnetization $\vec{M}_0$ is relatively thick, and the other (top) with free, 
precessing magnetization $\vec{M}$, is relatively thin. These layers are separated by a non-magnetic 
spacer. Characteristic widths of 100 nm allow to describe STO as a nano-device. If the current 
(in vertical direction) is applied, then when passing the fixed layer, the spin directions 
of the electrons align to the direction of $\vec{M}_0$. 
As these electrons enter the free layer, a spin 
transfer torque acts on its magnetization  $\vec{M}$, tending to reorient it, as
has been theoretically predicted by Slonczewski~\cite{Slonczewski-96} and Berger~\cite{Berger-96}. 
 As has been realized by  Slonczewski~\cite{Slonczewski-96}, 
 the spin transfer torque can compensate the 
damping of the spin precession of the free layer,
and in a constant external magnetic field a sustained oscillation (rotation of vector $\vec{M}$)
is observed.

After experimental observation of the generation~\cite{Kiselev_etal-03,Rippard_etal-04},
a lot of attention has been recently attracted to  synchronization of STOs. Indeed, as self-sustained
oscillators like electronic generators and lasers, they must demonstrate typical for this class of
physical systems effects of phase locking by external injection, and of
mutual synchronization if two or more
devices are coupled~\cite{Pikovsky-Rosenblum-Kurths-01}. Beside from the  fundamental
interest, synchronization of STOs is also of high practical relevance, as a way
to  increase the output power of otherwise rather weak individual STOs~\cite{PhysRevLett.101.017201}.

In the context of uniform STOs, the mostly promising 
way of coupling the STOs to achieve synchrony is to connect them in serial
electrically via
the common microwave 
current~\cite{Grollier-Cros-Fert-06,georges:232504,PhysRevB.82.140407,PhysRevB.84.104414,PhysRevB.86.014418}
(in experiments \cite{Kaka_etal-05,Mancoff_etal-05} a synchronization 
of two STOs was observed; however the coupling was not the electrical one, 
but due to spin waves, as the distance between two STOs build on the same mesa was about
500 nm).
In Ref.~\cite{Grollier-Cros-Fert-06} a 
prototype model for such a coupling has been suggested, where 
$N$ STOs are connected in series and are subject to a common dc current, with a parallel 
resistive load. The coupling is due to the giant magnetic resistance (GMR) effect, 
as the resistance of an STO $R_i$
depends on the orientation of its magnetization $\vec{M}_i$, so that the redistribution 
of the ac current between the STO array and the load 
depends on the the average (over the ensemble of $N$ STOs) value of this 
resistance $\langle R_i\rangle$. This situation is a typical mean-field coupling of oscillators, mostly prominent
examplified by the Kuramoto model~\cite{Kuramoto-84,Acebron-etal-05}. 
This setup has been further studied in 
\cite{PhysRevB.84.104414,PhysRevB.86.014418}, 
with a  more emphasis on nonlinear dynamical description of the ensemble behavior. 
The result of these studies is that synchronization is very hard to achieve, 
and if it is observed, it is
rather sensitive and not robust. 
 Also further numerical 
simulations~\cite{PhysRevB.82.140407,PhysRevB.84.104414} have shown a large variety of multistable
regimes including non-synchronized states. These observations have been 
recently confirmed by the analysis of
two coupled STOs~\cite{PhysRevB.86.014418}.

Another approach to study synchronization properties of STOs,  based not on the exact 
microscopic equations (Eqs.~(\ref{eq:llgs}) below), but on general equations for self-sustained 
oscillators has been proposed in \cite{PhysRevB.74.104401,Slavin-Tiberkevich-09} and followed
in~\cite{georges:232504,PhysRevB.86.104418,PhysRevB.82.012408}.  Because here the resulting 
dynamics is only suggested based on general qualitative arguments, but not derived from the microscopic equations,
predictions for synchronization properties do not 
extend beyond standard qualitative ones. In particular, in this approach one 
writes an effective Kuramoto-type
model for many mutually coupled STOs~\cite{georges:232504}, 
which does not represent the sensitivity
observed in the simulation based on the microscopic equations.

In this paper we study theoretically a serial array of STOs  subject to a dc current, with a
general  parallel 
inductor-capacitor-resistor (LCR) load. This setup is highly motivated by similar studies of synchronization of Josephson 
junctions~\cite{Wiesenfeld-Swift-95,Wiesenfeld-Colet-Strogatz-98}. 
In particular, in \cite{Filatrella_etall-00} such a model has been directly compared to the
experiments with Josephson junctions in a strongly resonant cavity~\cite{Barbara_etal-99}.
The LCR load can operate, depending on the frequency, either as an inductive one, or as a capacitive one. This flexibility allows one to find, similar to the case of Josephson junctions, situations with a robust synchronization of STOs.
We show also, that transition to synchrony as the parameter (dc current) varies, occurs through rather
complex states of partial synchrony and clustering, not presented in the standard Kuramoto model. 

\section{Basic model}

\begin{figure}[htb]
\centering
\includegraphics[width=0.2\textwidth]{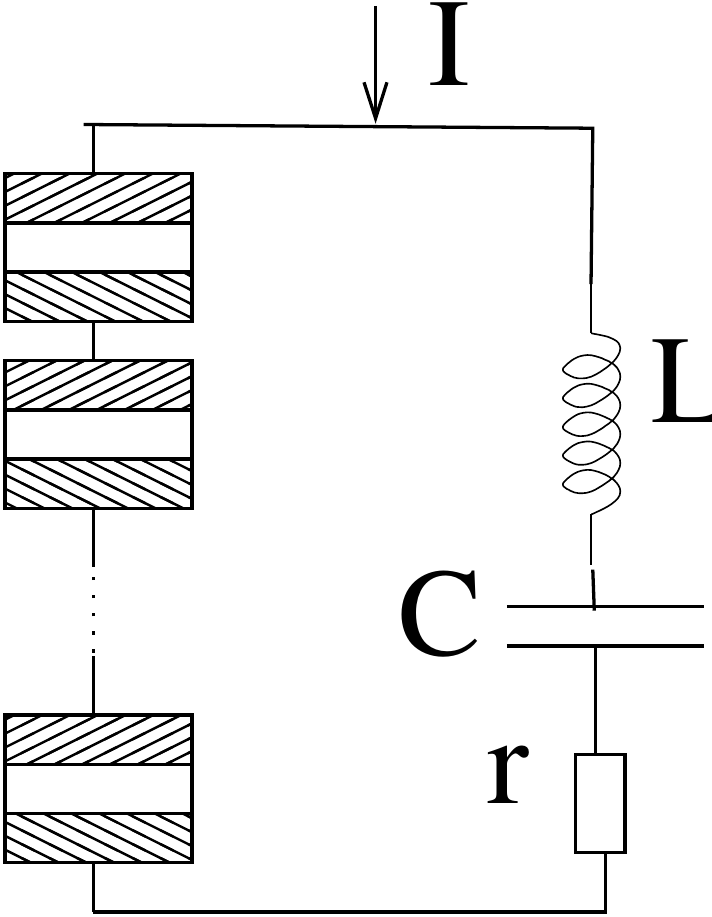}
\caption{The equivalent curcuit of the serial array of STO oscillators with a LCR load.}
\label{fig:circuit}
\end{figure}
 
 We consider an array of STOs with a LCR load as depicted in Fig.~\ref{fig:circuit}. The 
equations for the load are
\begin{equation}
LC\frac{d^2 V}{dt^2}+rC\frac{dV}{dt}+V=R(I-C\frac{dV}{dt})\;,
\label{eq:circ}
\end{equation}
where $R$ is the time-dependent resistance 
of the STO array, which is subject to current $J=I-C\frac{dV}{dt}$.
Eq.~(\ref{eq:circ}) is complemented by the system of equations for STOs.

Each STO is described by its free-layer magnetization $\vec{M}_i$ which obeys
the Landau-Lifshitz-Gilbert-Slonczewski equation
\begin{equation}
\frac{d}{dt}\vec{M}_i=-\gamma \vec{M}_i\times \vec{H}_{eff}+\alpha \vec{M}_i\times 
\frac{d}{dt}\vec{M}_i +\gamma\beta J \vec{M}_i\times(\vec{M}_i\times\vec{M}_0)\;,
\label{eq:llgs}
\end{equation}
where $\gamma$ is the gyromagnetic ratio; $\alpha$ is the Gilbert damping constant; 
$\beta$ contains material parameters; $J$ is the current through the STO; the effective 
magnetic field $H_{eff}$ contains an external magnetic field, an easy-axis field, 
and an easy-plane anisotropy field; $\vec{M}_0$ is magnetization of the fixed layer.

Following~\cite{PhysRevB.82.140407} we assume that
$\vec{H}_{eff}=H_a\hat{e}_x+(H_kM_x\hat{e}_x-H_{dz}M_z\hat{e}_z)/|\vec{M}|$.
Then, in spherical coordinates ($\phi,\theta$) the LLGS equations read~\cite{PhysRevB.82.140407}
\begin{equation}
\begin{aligned}
\frac{1+\alpha^2}{\gamma}\dot\theta_i&=U\cos\theta_i\cos\phi_i-W\sin\phi_i+\alpha S-T\;,\\
\frac{1+\alpha^2}{\gamma}\sin\theta_i 
\dot\phi_i&=-U\sin\phi_i-W\cos\phi_i\cos\theta_i-S-\alpha T\;,
\end{aligned}\label{eq:llgs-ph}
\end{equation}
where 
\begin{equation}
\begin{gathered}
S=(H_{dz}+H_k\cos^2\phi_i)\sin\theta_i\cos\theta_i,\qquad T=H_k\sin\phi_i\cos\phi_i\sin\theta_i\;,\\
U=\alpha H_a-\beta J\;,\qquad W=H_a+\alpha\beta J\;.
\end{gathered}
\label{eq:llgs-phpar}
\end{equation}

The system is closed by relating the resistance of the array $R$ to the states of
the STOs $\phi_i,\theta_i$. 
According to Ref.~\cite{Grollier-Cros-Fert-06}, the resistance depends on the angle 
$\delta$ between the magnetizations in the fixed and the free layers. In our case the magnetization 
of the fixed layer is along $x$-axis, therefore $\cos\delta=\sin\theta\cos\phi$. It is assumed that 
the resistance varies between value $R_P$ (parallel maagnetizations, $\delta=0$) and
$R_{AP}$ (antiparallel magnetizations, $\delta=\pi$) according to
\[
F(\theta,\phi) =\frac{R_P+R_{AP}}{2}-\frac{R_{AP}-R_P}{2}\cos\delta=R_0-R_1\sin\theta\cos\phi\;,
\] 
where $R_0=\frac{R_P+R_{AP}}{2}$ and $R_1=\frac{R_{AP}-R_P}{2}$.
Then we calculate $R$:
\begin{equation}
R=\sum_1^N (R_0-R_1\sin\theta\cos\phi)=\rho(1-\e X)\;,
\label{eq:r1}
\end{equation}
where
\begin{equation}
\rho=NR_0\;,\qquad\e=\frac{R_1}{R_0}\;,\qquad X=\langle\sin\theta\cos\phi\rangle=\frac{1}{N}\sum_1^N\sin\theta_i\cos\phi_i\;.
\label{eq:r2}
\end{equation}
The final system of equations is a combination of Eqs.~(\ref{eq:circ},\ref{eq:llgs-ph},\ref{eq:llgs-phpar},\ref{eq:r1},\ref{eq:r2}). We write it the dimensionless form, for the derivation we refer to the Appendix~\ref{sec:ap1}:
\begin{equation}
\begin{aligned}
\frac{d\theta_i}{dt}&=U\cos\theta_i\cos\phi_i-W\sin\phi_i+\alpha S-T\;,\\
\sin\theta_i 
\frac{d\phi_i}{dt}&=-U\sin\phi_i-W\cos\phi_i\cos\theta_i-S-\alpha T\;,\\
\frac{du}{dt}&=\frac{\omega}{N}w\;,\\
\frac{dw}{dt}&=
\frac{N\Omega^2}{\omega} [(1-\e X)(1-w)-u]\;,\\
&S=(H_{dz}+H_k\cos^2\phi_i)\sin\theta_i\cos\theta_i\;,\qquad T=H_k\sin\phi_i\cos\phi_i\sin\theta_i\;,\\
&U=\alpha H_a-\beta I(1-w)\;,\qquad W=H_a+\alpha\beta I(1-w)\;,\\
&X=\frac{1}{N}\sum_1^N\sin\theta_i\cos\phi_i\;.
\end{aligned}
\label{eq:bas}
\end{equation}
Here variables $\theta_i,\phi_i$ describe individual STOs in the array, $u\sim V$ and $w\sim \frac{dV}{dt}$ are global variables describing the load, and the interaction between these systems is via the mean field $X$.

Below we fix parameters of the STOs following Ref.~\cite{PhysRevB.82.140407} 
\[
H_a=0.2\;,\quad\alpha=0.01\;,\quad\beta=\frac{10}{3}\;,\quad H_{dz}=1.6\;,\quad H_k=0.05\;,
\]
focusing on the dependence on the dimensionless parameters
of the load $\Omega,\omega$, the coupling parameter $\e$,
and on the external current $I$.

\section{Dynamical states}

In the ensembles of globally coupled identical oscillators, different dynamical regimes
are generally possible:
\begin{itemize}
\item Complete synchrony, where states of all oscillators coincide, and coincide also with the mean 
fields. The dynamics reduces then to a low-dimensional system that includes the oscillator 
variables and the global fields (in our case the variable of the load). This regime is the 
mostly interesting one from the applied viewpoint, as here all the individual fields are summed coherently and the output field is maximal. It is, however, mostly boring from the dynamical viewpoint.
\item Clustered state, where oscillators form several clusters, within each of them their states 
coincide. Complete synchrony can be considered as the 1-cluster state; 
typically prevail states with a 
small number of clusters, but sometimes several clusters coexist with dispersed oscillators not belonging to clusters. Clustering means strong reduction of the number of independent variables, but the resulting regime can be quite complex as the dimension of the total system is larger
than one in the case of complete synchrony.
\item Asynchronous state, where all the oscillators remain different, and the mean field that 
mediates the interaction vanishes: one has in fact an ensemble of practically non-interacting
elements.
\item Partial synchronization, where the oscillators remain different but the mean fields do not vanish, and have macroscopic (compared with the finite-size fluctuations) values. 
These regimes are mostly difficult to describe, and a good theory exists in exceptional cases only~\cite{Mohanty-Politi-06,Pikovsky-Rosenblum-09}. 
The dynamics of the mean fields may be 
periodic~\cite{vanVreeswijk-96,Mohanty-Politi-06,Pikovsky-Rosenblum-09} or 
chaotic~\cite{Hakim-Rappel-92,Nakagawa-Kuramoto-94}.    
\end{itemize}

\begin{figure}[!htb]
\centering
\includegraphics[width=0.6\textwidth]{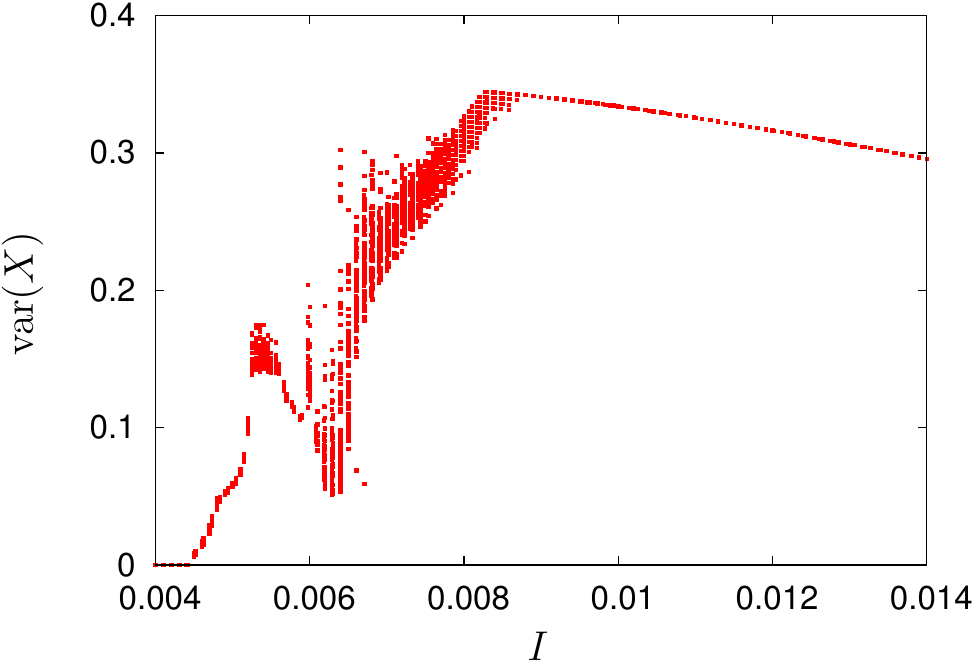}
\caption{Variation of the mean field 
for $\Omega=0.5$, $\e=0.3$, $\omega=1$, $N=200$ in dependence on the current $I$.
The time averaging has been performed over time interval $10^5$.}
\label{fig:mfI}
\end{figure}

Remarkably, for the considered array of STOs, we observe all these possible 
states, as described below. In Fig.~\ref{fig:mfI} we show the dependence of
the averaged over the time variation mean field 
$\text{var}(X)=\overline{(X(t)-\overline{X})^2}$ in dependence on the 
external current $I$, for $\omega=1$, $\Omega=0.5$, $\e=0.3$, $N=200$. To check for a
possible
multistability, for each set of parameters different runs starting from random 
initial conditions have been performed, and the values in each runs are shown with a marker.
Thus vertical spreading of markers indicates for  multistability; it is mostly pronounced 
for $0.006\lesssim I\lesssim 0.0085$, here clusters with different distributions are formed. For 
$0.0085\lesssim I$ one observes a complete synchrony, for $I\lesssim 0.0045$  an asynchronous state 
occurs, and for  $0.0045\lesssim I\lesssim 0.006$ a partial synchrony is observed. We describe these 
regimes in details below, using the same parameters as in Fig.~\ref{fig:mfI}.

\subsection{Complete synchrony and its stability}
\label{sec:css}

For a set of identical oscillators, the fully synchronous state usually means that 
all the dynamical variables of the array coincide. Here, because of symmetry $\theta,\phi\to\pi-\theta,-\phi$, the variables $\theta_i,\phi_i$ do not necessarily coincide, 
therefore we introduce
observables $x_i=\sin\theta_i\cos\phi_i$, $y_i=\cos\theta_i\sin\phi_i$ that are not affected by
the symmetry transformation. An additional advantage is that the mean field is just the average of $x_i$:
$
X=\frac{1}{N}\sum x_i
$.
Unfortunately, rewriting equations in these variables appears not possible, so we use them as ``observables'' to illustrate the dynamics, while performing calculations in the variables $\theta,\phi$.

\begin{figure}[!htb]
\centering
\includegraphics[width=\textwidth]{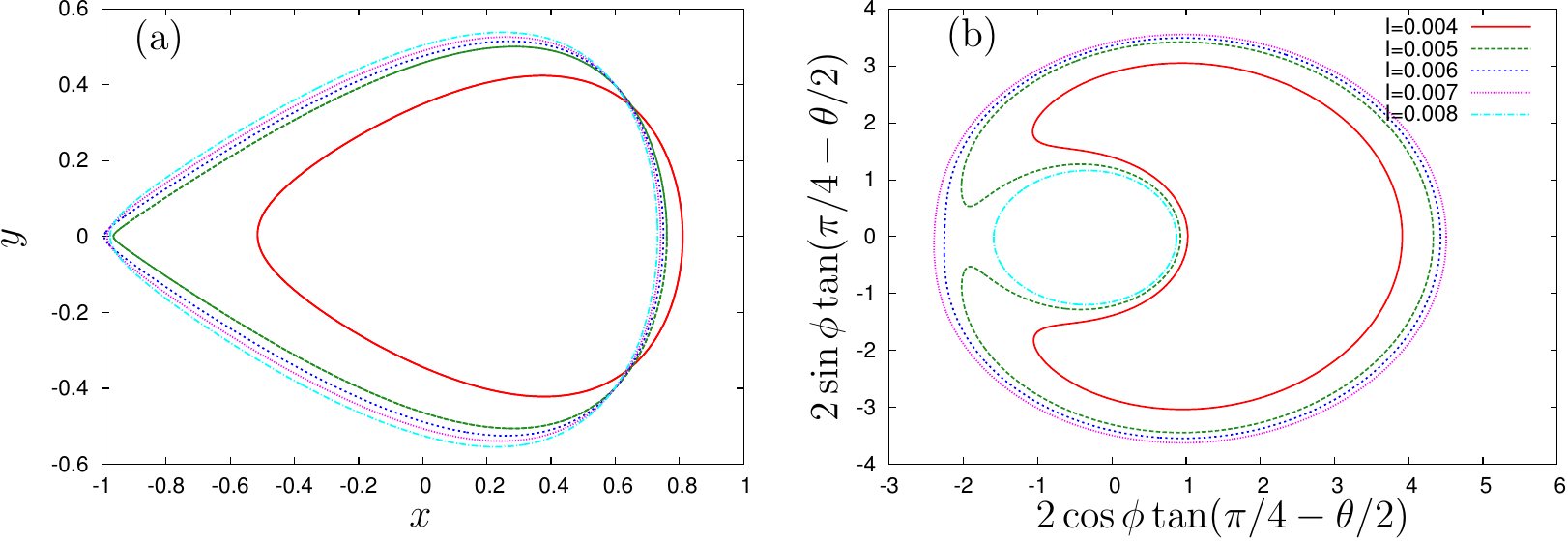}
\caption{(color online) Synchronous oscillations for $\Omega=0.5$, $\e=0.3$ $\omega=1$ and different $I$. (a) in coordinates $x,y$; (b) in the stereographic projection.}
\label{fig:cycle}
\end{figure}

In the fully synchronous regime system (\ref{eq:bas}) is a four-dimensional dissipative driven system of ODEs which in a large range of parameters possess a stable periodic (with period $T$) solution ($\theta^{0}(t),\phi^{0}(t), u^{0}(t),w^{0}(t)$) describing
STO oscillations. Depending on parameters, this limit cycle passes through 
a homoclinic bifurcation, at which its period becomes infinite, and the topology of the cycle on the sphere ($\phi,\theta$) changes (which is clearly seen in Fig.~\ref{fig:cycle}(b) as the transition from a ``small'' to a ``large'' cycle). We illustrate this in Fig.~\ref{fig:cycle}, where we show the cycle in the introduced coordinates $x,y$, and also in the stereographic projection.

Our next goal  is to establish stability of this periodic regime with respect to synchrony breaking. Starting with the synchronous solution of system (\ref{eq:bas}), we look
what happens if just one element of the array slightly deviates from this solution. This means 
that we perturb $\theta\to\theta^{0}+\delta\theta$, $\phi\to\phi^{0}+\delta\phi$, while keeping the 
mean field $X$ and the global variables $u,w$ at their values on the limit cycle (we do not write 
index at $\theta,\phi$ here because any oscillator can be perturbed, equations for the variations 
do not depend on the index). As a result, we get a two-dimensional linear system for $( \delta\phi,\delta\theta)$ with $T$-periodic coefficients determined by ($\theta^{0}(t),\phi^{0}(t), u^{0}(t),w^{0}(t)$). Solutions of such a Mathieu-type system are functions $\Theta_{1,2}(t)\exp(\lambda_{1,2} t),\Phi_{1,2}(t) \exp(\lambda_{1,2} t)$ where $\Theta_{1,2}(t)=\Theta_{1,2}(t+T), \Phi_{1,2}(t)=\Phi_{1,2}(t+T)$ are $T$-periodic. The resulting stability is determined
by multipliers $\mu_{1,2}=\exp[\lambda_{1,2}T]$: the perturbation $ ( \delta\phi,\delta\theta)$
decays if $|\mu_{1,2}|<1$ and increases otherwise. The calculation of this evaporation multiplier
\cite{Pikovsky-Popovych-Maistrenko-01} thus allows us to characterize linear stability of the synchronous cluster. 

\begin{figure}[!htb]
\centering
\includegraphics[width=0.7\textwidth]{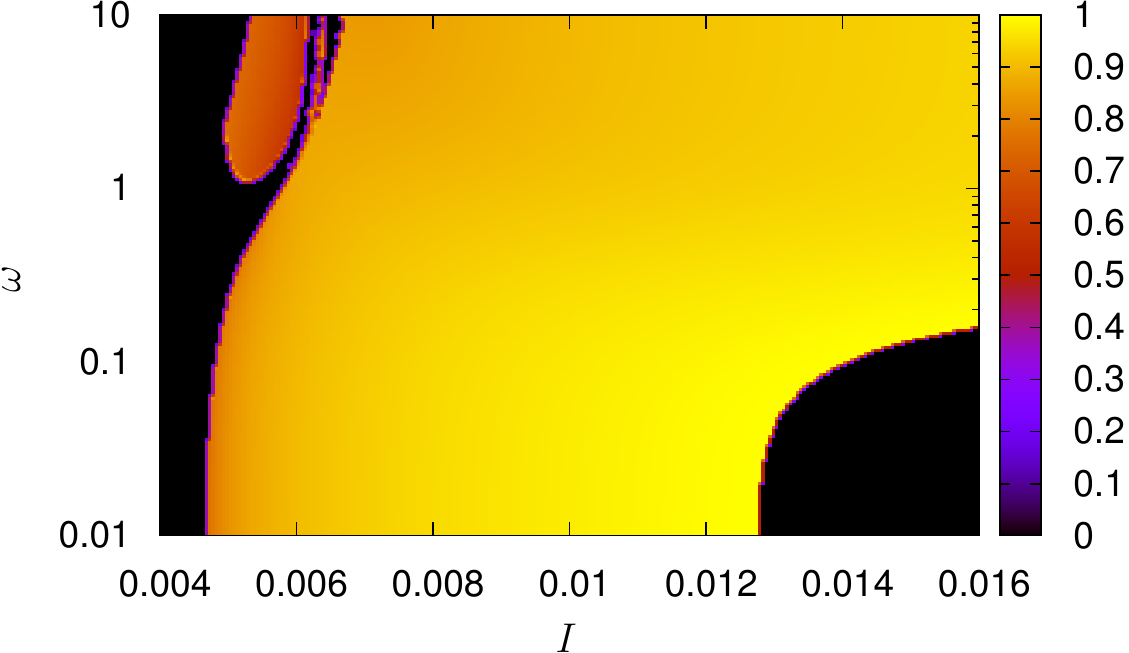}
\caption{(color online) Stability region for $\Omega=0.5$, $\e=0.3$: Instability is shown by black; otherwise color represents the absolute value of the largest evaporation multiplier.}
\label{fig:stab}
\end{figure}

The calculation of the evaporation multipliers is a straightforward numerical task after the 
periodic solution ($\theta^{0}(t),\phi^{0}(t), u^{0}(t),w^{0}(t)$) is found, we illustrate 
in Fig.~\ref{fig:stab} the stability region on plane of parameters ($\omega,I$). The smaller values of the largest evaporation multiplier correspond to stronger stability of synchrony and to more robust synchronization. In numerical simulations for $\omega=1$, we observed that the complete synchrony establishes for $0.009\lesssim I$: in all performed runs with $N=200$ starting from random initial conditions, either the full synchrony was established, or at the end of calculations a large cluster with almost all synchronous oscillators was observed, plus at most one or two that still did not belong to this majority cluster. This parameter range can be thus characterized as that of robust complete synchrony. For $I\lesssim 0.009$, although there is a region where the complete synchrony is stable, it does not typically evolve from the random initial conditions, as outlined below.

\subsection{Clusters}

\begin{figure}[!htb]
\centering
\includegraphics[width=0.7\textwidth]{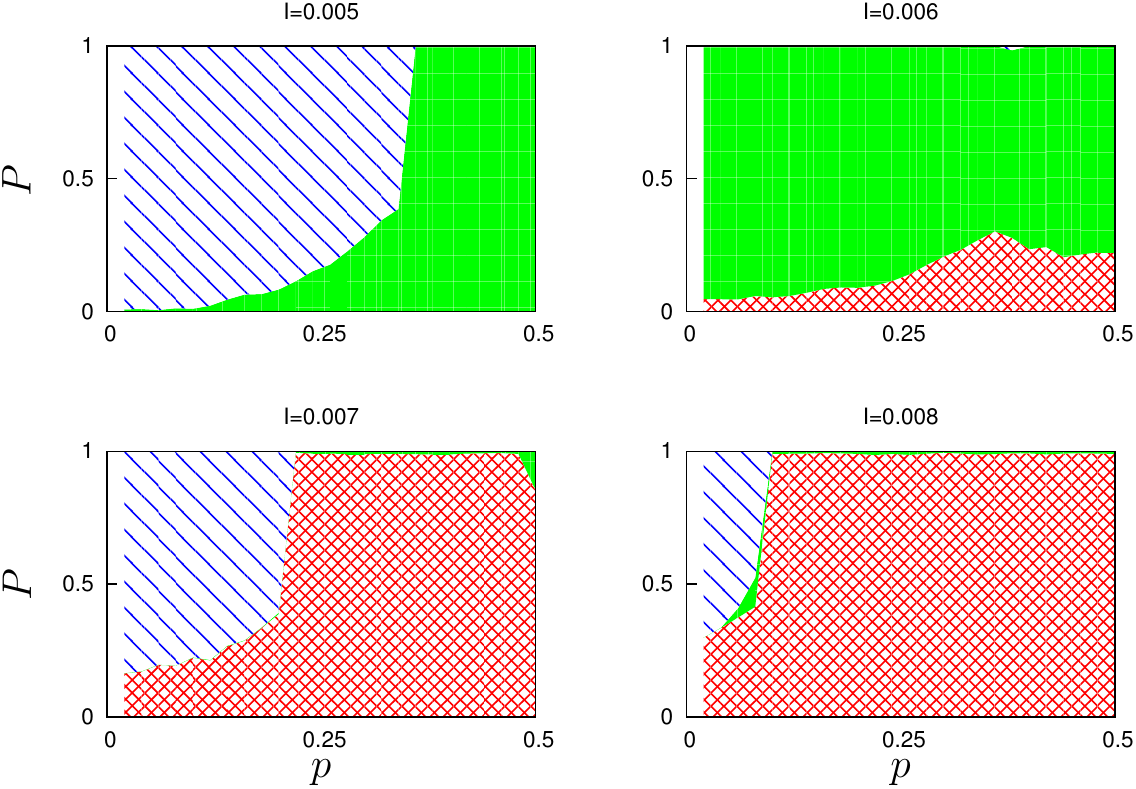}
\caption{(color online) Statistical evaluation of the evolution of initially randomly initialized 2-cluster state: propbability $P$ to observe different states vs. cluster distribution $p$, for several values of parameter $I$. Patterns from bottom to top depict regimes of one cluster (red, crossed pattern), 2 clusters with periodic dynamics (green, filled pattern), and 2 clusters with complex dynamics (mostly quasiperiodic, blue, inclined lines).}
\label{fig:cl2}
\end{figure}

The systematic study of all possible cluster states is hardly possible, so we restricted our attention
to two types of numerical experiments. In the first one, a statistical analysis of possible cluster 
states in the array of STOs has been performed. The simulations in the array
of $200$ oscillators with $\Omega=0.5,\e=0.3,\omega=1$ have been 
started from random initial conditions, and the final state after the
transient $t=10^{5}$ 
has been analyzed. For $0.004\leq I\leq 0.005$ no formation of large clusters have been observed. 
For $I=0.004$ typically all oscillators remain different, in few cases a small number of clusters of 
size 2 is built. For   $I=0.005$ building of many small clusters is typical, and clusters with 
sizes up to 
37 have been observed. For $I=0.006$ typically two-cluster states develop, but in $0.4\%$ of all runs
no essential clustering have been observed. For larger currents, $I\geq 0.007$, clustering was always observed. For $0.007\leq I\leq 0.008$ in many cases 
it was not complete: together with a large cluster, a set of non-clustered oscillators exists at the end of the transient time; for  $0.009\leq I\leq 0.014$ clustering was always a full one, with typically all oscillators fully synchronized (1-cluster state).

\begin{figure}[!htb]
\centering
\includegraphics[width=\textwidth]{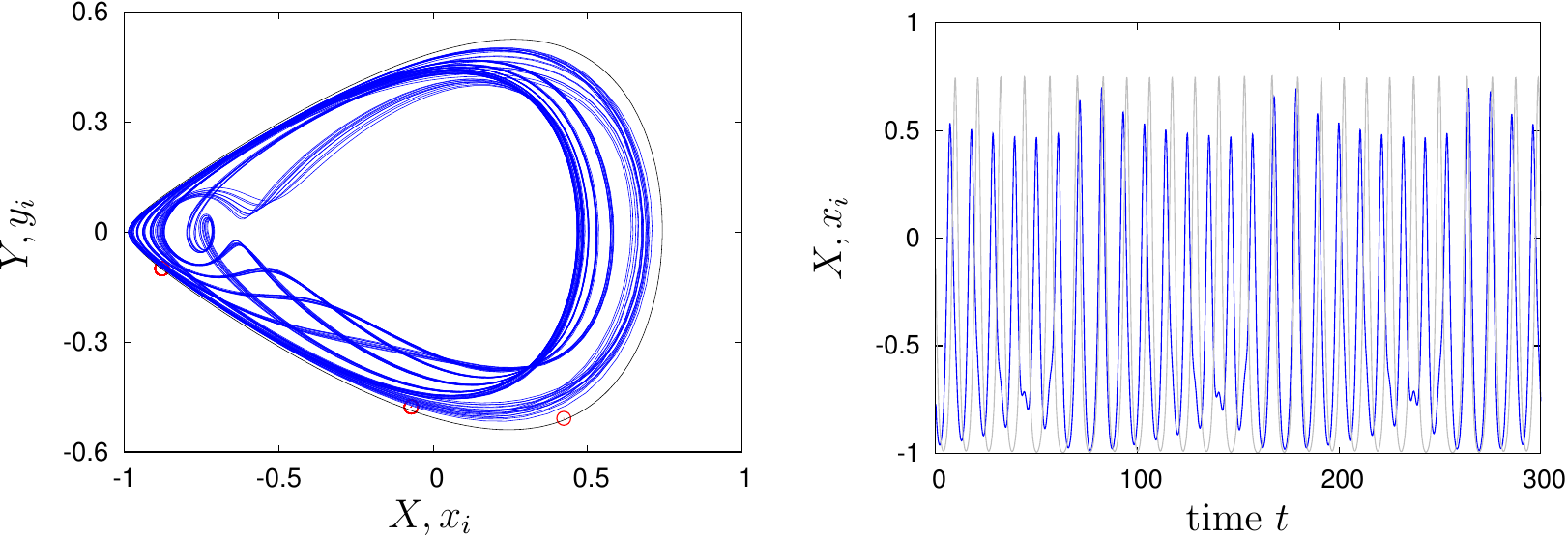}
\caption{(color online) A clustered regime for $I=0.007$. The array with $N=200$
 builds 2 clusters with $N_1=168$, $N_2=31$ and one separated oscillator. Left panel:
 a cluster state (red circles) and the evolution of the mean field (blue curve) in coordinates $X,x_i$ vs $Y,y_i$ . Black line shows for comparison the fully synchronous regime for these parameters.
 Right panel: the mean field $X(t)$ (blue curve) and the oscillations of one of the oscillators (grey curve). }
\label{fig:clusqp}
\end{figure}

In the second numerical experiment we initially prepared a 2-cluster state, with a given 
distribution $p$ between clusters ($p=N_1/N$, where $N_1$ is the number of oscillators in the first cluster). This 2-cluster state has been followed in time to see, if the clusters remain separated 
or they  merge to 1-cluster (complete synchrony). The results are shown in Fig.~\ref{fig:cl2}. In the range $0.009\leq I \leq 0.014$ practically all initial configurations eventually resulted in complete synchrony (not shown). For $I=0.004,0.005$ no one merging event has been observed; this corresponds to the fact that 1-cluster state is unstable for these parameters, as described above. For $0.006\leq I\leq 0.008$ both 2-cluster and 1-cluster states are observed.
For $I=0.007,0.008$ the 2-cluster states prevail for small $p$, i.e. for very asymmetric distribution among the clusters, here the regime is typically quasiperiodic. For $I=0.006$ the 2-cluster state is usually periodic, and there is a finite probability to merge. We illustrate a cluster state with a quasiperiodic mean field in Fig.~\ref{fig:clusqp}.

\subsection{Asynchronous state}
\begin{figure}[!htb]
\centering
\includegraphics[width=\textwidth]{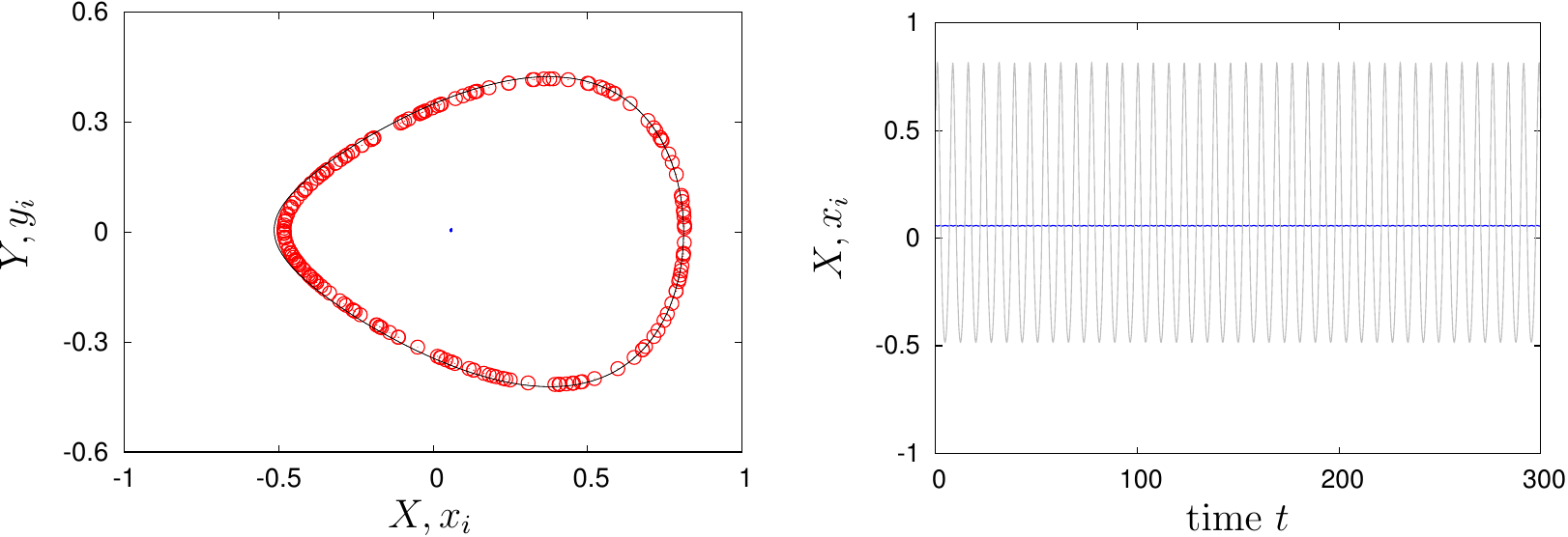}
\caption{(color online) An asynchronous regime for $I=0.004$.  Left panel:
 snapshot (red circles) of the state and the evolution of the mean field (blue curve, hardly seen at the origin) in coordinates $X,x_i$ vs $Y,y_i$. Black line shows for comparison the fully synchronous regime for these parameters.
 Right panel: the mean field $X(t)$ (blue curve) and the oscillations of one of the oscillators (grey curve).}
\label{fig:asyn}
\end{figure}

In this state, which is observed for $I\lesssim 0.0045$, the oscillators are uniformly (in time) distributed over the limit cycle, while the mean field vanishes. We illustrate this regime in Fig.~\ref{fig:asyn}.

\subsection{Partial synchrony}
\begin{figure}[!htb]
\centering
\includegraphics[width=\textwidth]{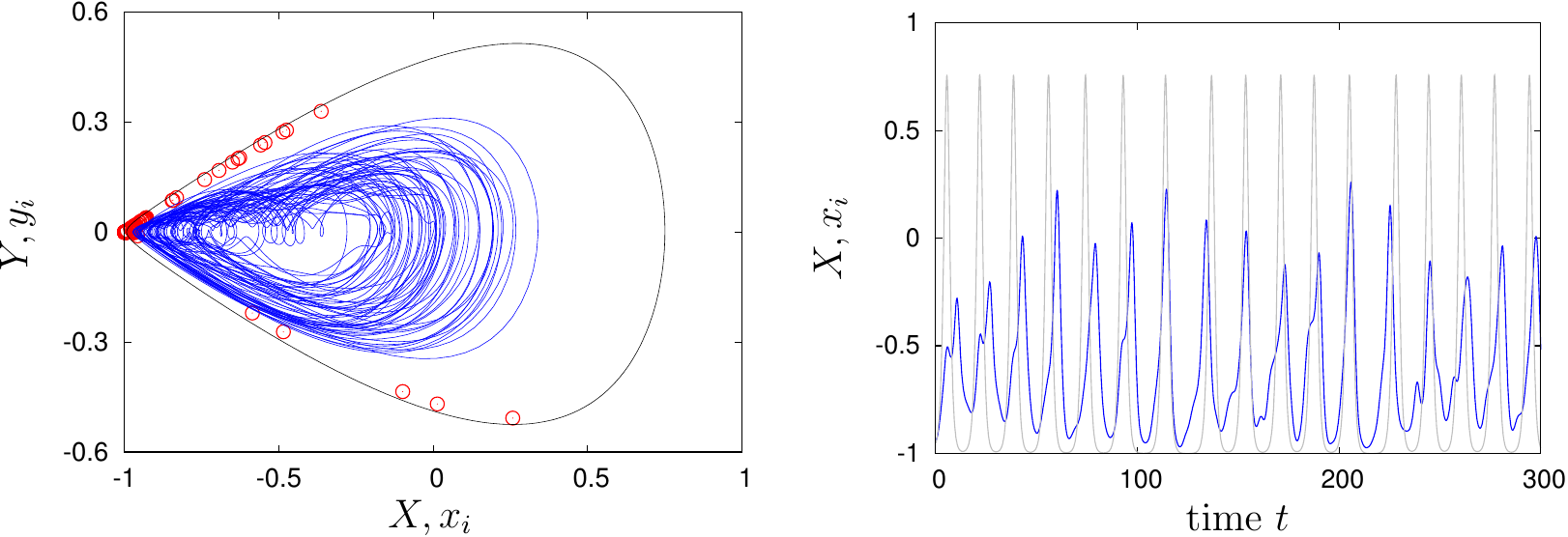}
\caption{(color online) A partially synchronous regime with chaotic mean field for $I=0.006$.  Left panel:
 snapshot (red circles) of the state and the evolution of the mean field (blue curve) in coordinates $X,x_i$ vs $Y,y_i$. Black line shows for comparison the fully synchronous regime for these parameters.
 Right panel: the mean field 
 $X(t)$ (blue curve) and the oscillations of one of the oscillators (grey curve).}
\label{fig:psyn}
\end{figure}

Regimes of partial synchrony, with a large number of clusters and non-vanishing mean field (which is nevertheless definitely smaller than in the case of full synchrony) are observed in the raange $0.0045\lesssim I\lesssim 0.006$. Typically these states are chaotic, as illustrated in Fig.~\ref{fig:psyn}. However, one cannot exclude that such a state is in fact a very long transient, and asymptotically for long times the clusters will ``grow''.
 
\section{Conclusion}
In this paper we have considered an array of spin-torque oscillators with a parallel 
inductor-capacitor-resistor load, and demonstrated that in this setup a robust synchronization
regime is observed in a wide range of parameters. While the region of stability of the synchronous regime is large, not always one observes full synchrony inside this domain: often oscillators 
organize themselves in several clusters, so that the synchrony is only partial. Transition from
asynchronous state to partial and full synchrony is rather nontrivial, with chaotic and quasiperiodic regimes of the mean field. These particular properties of the STO oscillators make an application of
simple models like that of Kuramoto model of phase coupled oscillators questionable. 

We have focused on the simplest setup, where all the oscillators are identical. To make realistic predictions, one has to take into account diversity of oscillator parameters (and, possibly, fluctuations); however the range of parameters (especially of that of the load) mostly promising for maximal synchrony can be estimated from the study of identical oscillators, as the region of maximal stability of the synchronous state and of absence of clustering. 
Remarkably, this stability can be determined in a rather simple way, by calculating
the evaporation exponent of the synchronous 1-cluster state as described in Sect.~\ref{sec:css}.
One can easily adopt this method to other sets of parameters, and to other types of load.
Another possible extension of this study is consideration of coupling schemes beyond the globally coupled ones, similar to the corresponding studies
of one-dimensional Josephson junction arrays~\cite{Vasilic_etal-03}; here however methods developed for global coupling cannot be directly applied.

\begin{acknowledgments}
The authors thanks NEXT program for support, IRSAMC for hospitality, and
 D. Shepelyansky and M. Rosenblum for
helpful discussions. 
\end{acknowledgments}

 \appendix 
 \section{Equations in dimensionless form}
 \label{sec:ap1}
We combine together  Eqs.~(\ref{eq:circ},\ref{eq:llgs-ph},\ref{eq:llgs-phpar},\ref{eq:r1},\ref{eq:r2}):
\begin{equation}
\begin{gathered}
\frac{1+\alpha^2}{\gamma}\dot\theta_i=U\cos\theta_i\cos\phi_i-W\sin\phi_i+\alpha S-T\\
\frac{1+\alpha^2}{\gamma}\sin\theta_i 
\dot\phi_i=-U\sin\phi_i-W\cos\phi_i\cos\theta_i-S-\alpha T\\
S=(H_{dz}+H_k\cos^2\phi_i)\sin\theta_i\cos\theta_i\quad T=H_k\sin\phi_i\cos\phi_i\sin\theta_i\\
U=\alpha H_a-\beta (I-C\frac{dV}{dt})\quad W=H_a+\alpha\beta (I-C\frac{dV}{dt})\\
LC\frac{d^2 V}{dt^2}+rC\frac{dV}{dt}+V=N R_0(1-\e X)(I-C\frac{dV}{dt})
\end{gathered}
\label{eq:llgs-ph1}
\end{equation}

We introduce new time $t'=\frac{\gamma}{1+\alpha^2}t$ and dimensionless voltage $v=\frac{V}{R_0I}$, and obtain, 
%
 denoting $\frac{\gamma^2}{(1+\alpha^2)^2}LC=\Omega^{-2}$ and $\frac{NR_0  C\gamma}{1+\alpha^2}=\omega^{-1}$ 

\begin{equation}
\begin{gathered}
\dot\theta_i=U\cos\theta_i\cos\phi_i-W\sin\phi_i+\alpha S-T\\
\sin\theta_i 
\dot\phi_i=-U\sin\phi_i-W\cos\phi_i\cos\theta_i-S-\alpha T\\
S=(H_{dz}+H_k\cos^2\phi_i)\sin\theta_i\cos\theta_i\quad T=H_k\sin\phi_i\cos\phi_i\sin\theta_i\\
U=\alpha H_a-\beta I(1-\frac{1}{N\omega}\frac{dv}{dt})\quad W=H_a+\alpha\beta I(1-\frac{1}{N\omega}\frac{dv}{dt})\\
\frac{1}{\Omega^2}\frac{d^2 v}{dt^2}+\frac{r}{R_0}\frac{1}{N\omega} \frac{dv}{dt}+v=N (1-\e X)(1-\frac{1}{N\omega}\frac{dv}{dt})
\end{gathered}
\label{eq:llgs-ph3}
\end{equation}

Finally, introducing $v=Nu$ and $w=\frac{1}{\omega}\frac{du}{dt}$ we obtain system (\ref{eq:bas}).
Additional parameters, related to the load, are: $\Omega$, $\omega$, and $r/(NR_0)$.
The latter parameter can be set to zero if the resistance of the load is much smaller 
than that of the STO array.

%



\end{document}